# Diffraction Problem and Amplitudes-Phases Dispersion of Eigen Fields of a Nonlinear Dielectric Layer

Vasyl V. Yatsyk[*]

Usikov Institute for Radiophysics & Electronics
National Academy of Sciences of Ukraine

## Abstract

The open nonlinear electrodynamic system - nonlinear transverse non-homogeneous dielectric layer, is an example of inorganic system having the properties of self-organization, peculiar to biological systems. The necessary precondition of effects of self-organization is the presence of a flow of energy acting in system from an external source, due to which the system gets ability to independent formation of structures. On an example of the transverse non-homogeneous, isotropic, nonmagnetic, linearly polarized, nonlinear (a Kerr-like dielectric nonlinearity) dielectric layer the constructive approach of the analysis of amplitudes-phases dispersion of eigen oscillation-wave fields of nonlinear object are shown. The norm of an eigen field is defined from the solution of a diffraction problem of plane waves or excitation of point or compact source of a nonlinear layer.

## 1 The Equations of the Electromagnetic Field

Maxwell's equations:

$$\nabla \times \vec{E}(\vec{r},t) = -\frac{1}{c}\frac{\partial \vec{B}(\vec{r},t)}{\partial t}, \quad \nabla \times \vec{H}(\vec{r},t) = \frac{1}{c}\frac{\partial \vec{D}(\vec{r},t)}{\partial t},$$
$$\nabla \cdot \vec{D}(\vec{r},t) = 0, \qquad \nabla \cdot \vec{B}(\vec{r},t) = 0, \tag{1}$$

and the material equations:

$$\vec{D}(\vec{r},t) = \vec{E}(\vec{r},t) + 4\pi\vec{P}(\vec{r},t),$$
$$\vec{B}(\vec{r},t) = \vec{H}(\vec{r},t) + 4\pi\vec{M}(\vec{r},t). \tag{2}$$

When $\vec{M}(\vec{r},t) = 0$ the equations (1), (2) are reduced to (see [1]):

$$\nabla^2 \vec{E}(\vec{r},t) - \nabla(\nabla\vec{E}(\vec{r},t)) - \frac{1}{c^2}\frac{\partial^2}{\partial t^2}\vec{D}^{(L)}(\vec{r},t) - \frac{4\pi}{c^2}\frac{\partial^2}{\partial t^2}\vec{P}^{(NL)}(\vec{r},t) = 0. \tag{3}$$

[*] E-mail address: yatsyk@vk.kharkov.ua, yatsyk@ire.kharkov.ua

Here: $\vec{D}^{(L)} = \vec{E} + 4\pi\vec{P}^{(L)} = \hat{\varepsilon}\vec{E}$; $\vec{P}^{(L)} = \hat{\chi}^{(1)}\vec{E}$; $\vec{D}_i^{(L)} = \varepsilon_{ij}^{(L)}\vec{E}_j$; $\varepsilon_{ij}^{(L)} = 1 + 4\pi\chi_{ij}^{(1)}$; $\vec{E}=(E_x,E_y,E_z)$; $\vec{P}=(P_x,P_y,P_z)$; $P_i \equiv P_i^{(L)} + P_i^{(NL)}$; $P_i^{(L)} \equiv \chi_{ij}^{(1)} E_j$; $P_i^{(NL)} \equiv \chi_{ijk}^{(2)} E_j E_k + \chi_{ijkl}^{(3)} E_j E_k E_l + \ldots$; $\varepsilon_{ij}^{(L)}$ are components of a tensor of a linear part of dielectric permittivity $\hat{\varepsilon}$; accordingly these parameters $\chi_{ij}^{(1)}$, $\chi_{ijk}^{(2)}$, $\chi_{ijkl}^{(3)}$, … are components of the appropriate tensors of susceptibilities $\hat{\chi}^{(1)}$, $\hat{\chi}^{(2)}$, $\hat{\chi}^{(3)}$, ….

Let $\vec{E}(\vec{r},t) = \exp(-i\omega t)\cdot \vec{E}(\vec{r})$. We consider a nonmagnetic $\vec{M}=0$, isotropic, transverse non-homogeneous $\varepsilon^{(L)}(z) = \varepsilon_{xx}^{(L)}(z)$, linearly polarized $\vec{E}=(E_x,0,0)$, $\vec{H}=(0,H_y,H_z)$ (E-polarized) and Kerr-like nonlinearity $P_x^{(NL)} = (3/4)\chi_{xxxx}^{(3)}|E_x|^2 E_x$ (where $\vec{P}^{(NL)} = (P_x^{(NL)}, 0, 0)$) dielectric layer (Fig. 1), [1-3].

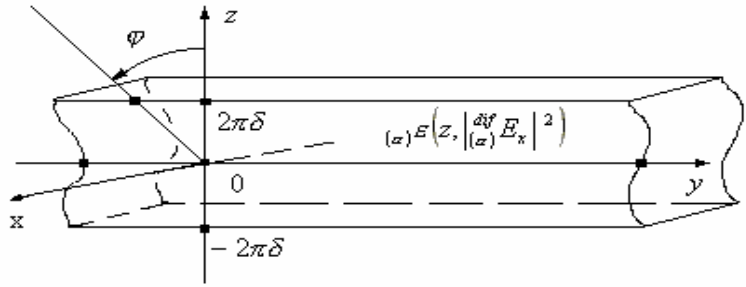

**Figure 1.** Nonlinear dielectric layer.

In this case (see (1), (2) and [1]): $\nabla\cdot\vec{D}=0 \Rightarrow \nabla\cdot\vec{E} = -(\vec{E}\cdot(\nabla\cdot\hat{\varepsilon})/\hat{\varepsilon}) \Rightarrow \nabla\cdot(\nabla\cdot\vec{E})=0$, and the equation (3) has the following kind:

$$\Delta\cdot\vec{E} + \frac{\omega^2}{c^2}\cdot\varepsilon^{(L)}(z)\cdot\vec{E} + \frac{4\pi\omega^2}{c^2}\cdot\vec{P}^{(NL)} \equiv$$
$$\equiv \left(\Delta + \kappa^2\cdot{}_{(\alpha)}\varepsilon(z,|E_x|^2)\right)\cdot E_x(y,z) = 0. \quad (4)$$

Here: ${}_{(\alpha)}\varepsilon(z,|E_x|^2) = \begin{cases} 1, & |z| > 2\pi\delta \\ \varepsilon^{(L)}(z) + \alpha\cdot|E_x|^2, & |z| \le 2\pi\delta \end{cases}$; $\Delta = \partial^2/\partial y^2 + \partial^2/\partial z^2$; $\alpha = 3\pi\chi_{xxxx}^{(3)}$.

## 2 The Spectral Problems for Nonlinear Dielectric Layer

### 2.1 Formulation of the Nonlinear Spectral Problems

The spectral problems for transverse non-homogeneous along an axis $0z$, homogeneous along an axis $0x$ and longitudinal direction $0y$, nonlinear dielectric layer (a Kerr-like nonlinearity) $\{(x,y,z): -\infty < x < \infty, -\infty < y < \infty, -2\pi\delta \le z \le 2\pi\delta\}$ of height $4\pi\delta$ and parameter of dielectric permeability ${}_{(\alpha)}\varepsilon(z,|{}_{(\alpha)}^{(nor)}E_x|^2)$ (case of E-polarization, ${}_{(\alpha)}^{(nor)}E_y = {}_{(\alpha)}^{(nor)}E_z = {}_{(\alpha)}^{(nor)}H_x = 0$, see Fig. 1 and (4)) is considered:

$$\left[\frac{\partial^2}{\partial y^2}+\frac{\partial^2}{\partial z^2}+\kappa^2{}_{(\alpha)}\varepsilon(z,|{}^{(nor)}_{(\alpha)}E_x|^2)\right]{}^{(nor)}_{(\alpha)}E_x(y,z)=0, \qquad (5)$$

the generalized boundary conditions:

$${}^{(nor)}_{(\alpha)}E_{tg} \text{ and } {}^{(nor)}_{(\alpha)}H_{tg} \text{ are continuous at discontinuities } {}_{(\alpha)}\varepsilon(z,|{}^{(nor)}_{(\alpha)}E_x|^2);$$
$${}^{(nor)}_{(\alpha)}E_x(y,z)={}^{(nor)}_{(\alpha)}U(z)\exp(i\phi y) \text{ - condition of spatial quasi homogeneity on } y; \qquad (6)$$

condition of the radiation

$${}^{(nor)}_{(\alpha)}E_x(y,z)=\begin{Bmatrix}{}^{(nor)}_{(\alpha)}a\\{}^{(nor)}_{(\alpha)}b\end{Bmatrix}\exp[i(\phi y \pm \Gamma\cdot(z \mp 2\pi\delta))], \quad z \begin{matrix}>\\<\end{matrix} \pm 2\pi\delta, \qquad (7)$$

where, specifically for real $\kappa$ and $\phi$,

$$\text{Im}\,\Gamma \geq 0,\ \text{Re}\,\Gamma\cdot\text{Re}\,\kappa \geq 0,\ \text{for Im}\,\phi = 0,\ \text{Im}\,\kappa = 0. \qquad (8)$$

Condition (4) are physical, since they imply that no waves arise from infinity $|z|=\infty$.

Here: ${}^{(nor)}_{(\alpha)}E_x$, ${}^{(nor)}_{(\alpha)}E_{tg}$, ${}^{(nor)}_{(\alpha)}H_{tg}$ - are the designated normalized component of fields of the vectors $\vec{E}$ and $\vec{H}$; $\Gamma=(\kappa^2-\phi^2)^{1/2}$; the index $(nor)$ specifies that are considered the normalized components of eigen fields, the index $(\alpha)$ means that corresponds value $\alpha = const$, where $\alpha$ - parameter cubic electric susceptibility, $\text{Re}\,\alpha \equiv const \geq 0$, $\text{Im}\,\alpha \equiv 0$;

$${}_{(\alpha)}\varepsilon\!\left(z,\left|{}^{(nor)}_{(\alpha)}E_x\right|^2\right)=\begin{cases}1,|z|>2\pi\delta\\ \varepsilon^{(L)}+\alpha\cdot\left|{}^{(nor)}_{(\alpha)}E_x\right|^2\cdot e^{2\text{Im}\phi\cdot y},|z|\leq 2\pi\delta\end{cases}\equiv\begin{cases}1,|z|>2\pi\delta\\ \varepsilon^{(L)}(z)+\alpha\cdot\left|{}^{(nor)}_{(\alpha)}U(z)\right|^2,|z|\leq 2\pi\delta\end{cases} \qquad (9)$$

- parameter of dielectric permeability of environment for nonlinear spectral problems (in (9) the multiplier $e^{2\text{Im}\phi\cdot y}$ compensates growth or decrease of dielectric permeability along a layer); $\varepsilon^{(L)}(z)$ - piecewise smooth function of variable $z$, $\text{Re}\,\varepsilon^{(L)}(z)>0$, $\text{Re}\,\kappa\,\text{Im}\,\varepsilon^{(L)}(z)\geq 0$; $\exp(-i\omega t)$ - temporary dependence; dimensionless parameters: $\omega=\kappa c$ - circular frequency; $\kappa=\omega/c\equiv 2\pi/\lambda$ - frequency, describing the attitude of true height $h$ of a layer to length $\lambda$ of a wave of excitation, $h/\lambda=2\kappa\delta$; $c=(\varepsilon_0\mu_0)^{-1/2}$, $\text{Im}\,c=0$, $\varepsilon_0$ and $\mu_0$ - material parameters of environment.

## 2.2 The Spectral Parameters on the Complex Riemannian Manifolds

A range of values of frequency spectral parameter $\kappa\in{}_{(\alpha)}\Omega_\kappa$, that of propagation eigen constant $\phi\in{}_{(\alpha)}\Omega_\phi$ and that of the generalized spectral parameter $(\kappa,\phi)\in{}_{(\alpha)}\Omega_{(\kappa,\phi)}$ represent the two-sheeted complex Riemannian manifold $H$, $\Phi$ and $H\times\Phi$, respectively (see [4-6]). Here ${}_{(\alpha)}\Omega_\kappa$, ${}_{(\alpha)}\Omega_\phi$ and ${}_{(\alpha)}\Omega_{(\kappa,\phi)}$ - spectral sets. They are fully identified by the boundaries of the regions of complex $\phi$ (for the surface $\Phi$), complex $\kappa$ (for the surface H) and complex $(\kappa,\phi)$ (for the manifold $H \times \Phi$) where, in accordance with (8), the canonical Green's function

$$G_0(y,z;y_0,z_0;\kappa,\phi) = \frac{i}{4\pi}\exp[i(\phi(y-y_0)+\Gamma\cdot|z-z_0|)]/\Gamma \qquad (10)$$

of unperturbed $(_{(\alpha)}\varepsilon(z,|_{(\alpha)}^{(nor)}E_x|^2) \equiv 1)$ problem (5)-(7) can be analytically continued in the absence of scattering objects. The manifolds H, $\Phi$ and $H \times \Phi$ consists of two sheets, the branch points ("branches") are located by the conditions $\kappa^{\pm}: \Gamma(\kappa^{\pm})=0$ ($\kappa^{\pm}=\pm|\phi|$, $\phi = const \in R$), $\phi^{\pm}: \Gamma(\phi^{\pm})=0$ ($\phi^{\pm}=\pm|\kappa|$, $\kappa = const \in R$), $\Gamma(\kappa^{\pm},\phi^{\pm})=0$ ($\kappa^{\pm}=\pm\phi^{\pm}$) in conjunction with the cuts issuing out of these points along the curves ("surfaces")

$$\begin{cases} (\mathrm{Re}\,\kappa)^2 - (\mathrm{Im}\,\kappa)^2 - (\mathrm{Re}\,\phi)^2 + (\mathrm{Im}\,\phi)^2 = 0, \\ \mathrm{Re}\,\kappa(\mathrm{Re}\,\kappa\mathrm{Im}\,\kappa - \mathrm{Im}\,\phi\mathrm{Re}\,\phi) \leq 0 \end{cases}. \qquad (11)$$

The first, "physical" sheet of the surfaces H, $\Phi$, and $H \times \Phi$ (the pairs $\{(\kappa,\phi),\Gamma(\kappa,\phi)\}$) is fully identified by radiation condition (8) and cuts (11). The second, "nonphysical" sheets of the surfaces H, $\Phi$, and $H \times \Phi$ differ from the "physical" ones by the change of signs at $\mathrm{Re}\,\Gamma$ and $\mathrm{Im}\,\Gamma$.

## 2.3 The Homogeneous Nonlinear Integral Equations and Untrivial Solutions

Eigen fields of free fluctuations, eigen waves, eigen oscillation-wave modes of a layer $_{(\alpha)}\varepsilon(z,|_{(\alpha)}^{(nor)}E_x|^2)$ adequate, accordingly, eigen meanings of frequencies $\kappa \in {}_{(\alpha)}\Omega_\kappa(\phi = const \in R) \subset H$, constant distribution $\phi \in {}_{(\alpha)}\Omega_\phi(\kappa = const \in R) \subset \Phi$, generalized meanings of spectra $(\kappa,\phi) \in {}_{(\alpha)}\Omega_{(\kappa,\phi)} \subset H \times \Phi$ accept as:

$$_{(\alpha)}^{(nor)}E_x(y,z) = \begin{cases} {}_{(\alpha)}^{(nor)}a\exp[i(\phi y + \Gamma\cdot(z - 2\pi\delta))], & z > 2\pi\delta, \\ {}_{(\alpha)}^{(nor)}U(z)\exp(i\phi y), & |z| \leq 2\pi\delta, \\ {}_{(\alpha)}^{(nor)}b\exp[i(\phi y - \Gamma\cdot(z + 2\pi\delta))], & z < -2\pi\delta. \end{cases} \qquad (12)$$

Where $_{(\alpha)}^{(nor)}U(-2\pi\delta) = {}_{(\alpha)}^{(nor)}b$, $_{(\alpha)}^{(nor)}U(2\pi\delta) = {}_{(\alpha)}^{(nor)}a$; $_{(\alpha)}\Omega_\kappa$, $_{(\alpha)}\Omega_\phi$, $_{(\alpha)}\Omega_{(\kappa,\phi)}$ - spectral sets, H, $\Phi$, $H \times \Phi$ - the two-sheeted complex Riemannian manifold; $R$ - set of real numbers.

As well as in [4-7] (similarly to results of the analysis of a linear spectral problem for a gratings, see [4]) the nonlinear spectral problems (5)-(7) are reduced to a finding of the untrivial normalized decisions $_{(\alpha)}^{(nor)}U(z) \in L_2([-2\pi\delta, 2\pi\delta])$ of the homogeneous nonlinear integrated equation second-kind

$$_{(\alpha)}^{(nor)}U(z) + \frac{i\kappa^2}{2\Gamma(\kappa,\phi)}\int_{-2\pi\delta}^{2\pi\delta} e^{i\Gamma(\kappa,\phi)|z-z_0|}\left[1 - \left(\varepsilon^{(L)}(z_0) + \alpha|_{(\alpha)}^{(nor)}U(z_0)|^2\right)\right]_{(\alpha)}^{(nor)}U(z_0)\,dz_0 = 0, |z| \leq 2\pi\delta, \quad (13)$$

designations here are used $_{(\alpha)}\varepsilon(z,|_{(\alpha)}^{(nor)}E_x|^2) \equiv \varepsilon^{(L)}(z) + \alpha|_{(\alpha)}^{(nor)}U(z)|^2$, $_{(\alpha)}^{(nor)}U(z) \equiv _{(\alpha)}^{(nor)}U(z,\kappa,\phi)$.

In [5] is given algorithms of the decision of nonlinear spectral problems (5)-(7) for the normalized own field (the problem of a norm of eigen fields of a nonlinear layer is considered below in section 4). It is based on the solution of an equivalent problem for the nonlinear homogeneous integrated equation (13). The solution (13) is with application of a quadrature method, is reduced to homogeneous system of the nonlinear equations of the second-kind with nonlinear entry of spectral parameter $\kappa$, $\phi$, $(\kappa,\phi)$.

The untrivial normalized solutions $_{(\alpha)}^{(nor)}U$ of a nonlinear problem (5)-(7) and appropriate eigen value $\kappa \in _{(\alpha)}\Omega_\kappa$, $\phi \in _{(\alpha)}\Omega_\phi$, $(\kappa,\phi) \in _{(\alpha)}\Omega_{(\kappa,\phi)}$ are determined by the solution of system equations consisting from the characteristic equation and nonlinear homogeneous system of the equations equivalent to the equation (13) (see [5])

$$\begin{cases} _{(\alpha)}f\left(\kappa,\phi,\left|_{(\alpha)}^{(nor)}U\right|^2\right) = \det\left(E - _{(\alpha)}B\left(\kappa,\phi,\left|_{(\alpha)}^{(nor)}U\right|^2\right)\right) = 0, \\ (E - _{(\alpha)}B(\kappa,\phi,\left|_{(\alpha)}^{(nor)}U\right|^2)) \cdot _{(\alpha)}^{(nor)}U = 0, \end{cases} \begin{cases} \kappa \in _{(\alpha)}\Omega_\kappa \ (\phi = const \in R) \subset H, \\ \phi \in _{(\alpha)}\Omega_\phi \ (\kappa = const \in R) \subset \Phi, \\ (\kappa,\phi) \in _{(\alpha)}\Omega_{(\kappa,\phi)} \subset H \times \Phi. \end{cases} \quad (14)$$

Here $_{(\alpha)}^{(nor)}U = \{_{(\alpha)}^{(nor)}U_n\}_{n=1}^N$ - vector-column of unknown $_{(\alpha)}^{(nor)}U_n = _{(\alpha)}^{(nor)}U(z_n,\kappa,\phi)$, given in units, $z_1 = -2\pi\delta < z_2 < ... < z_n < ... < z_N = 2\pi\delta$, $n = 1,2,...,N$; $N$ - number of units, determining the order of system (14); $E = \{\delta_{nm} \equiv \delta_n^m\}_{n,m=1}^N$ (where $\delta_n^m$ - Kronecker delta) and $_{(\alpha)}B\left(\kappa,\phi,\left|_{(\alpha)}^{(nor)}U\right|^2\right) = \left\{A_m \cdot _{(\alpha)}K_{nm}\left(\kappa,\phi,\left|_{(\alpha)}^{(nor)}U\right|^2\right)\right\}_{n,m=1}^N$ - matrix of dimension $N \times N$; $_{(\alpha)}K_{nm}\left(\kappa,\phi,\left|_{(\alpha)}^{(nor)}U\right|^2\right) = \frac{i\kappa^2}{2\Gamma(\kappa,\phi)} \cdot \exp(i\Gamma(\kappa,\phi) \cdot |z_n - z_m|) \cdot \left[1 - \left(\varepsilon^{(L)}(z_m) + \alpha \cdot \left|_{(\alpha)}^{(nor)}U_m\right|^2\right)\right]$; $A_m$ - numerical coefficients dictated by chosen quadrature form.

## 3 About Excitation of a Nonlinear Dielectric Layer

### 3.1 The Nonlinear Diffraction Problem

The amplitude-phase dispersion of the eigen oscillation-wave fields of nonlinear electrodynamic structure essentially depend on a source of excitation of nonlinear object [8], its amplitude, phase, spatial characteristics. As sources of excitation of electrodynamic structures usually use excitation by a plane wave, beam of plane waves [9], point or compact source [4]. The diffraction fields adequate to a volume or other source of excitation of nonlinear structure are determining at a choice of norm of an eigen field of a nonlinear spectral problem. A base problem, with which use the algorithms of the solution of each of the listed problems of excitation are under construction, is the diffraction problem of a plane wave on nonlinear electrodynamic object.

We shall consider a diffraction field (case of E-polarization) received at fall of a plane wave $^{inc}E_x(y,z) = ^{inc}a \cdot \exp[i(\phi y - \Gamma \cdot (z - 2\pi\delta))]$, $z > 2\pi\delta$ on a nonlinear layer $_{(\alpha)}\varepsilon\left(z,\left|_{(\alpha)}^{dif}E_x\right|^2\right) = \begin{cases} 1, & |z| > 2\pi\delta \\ \varepsilon^{(L)}(z) + \alpha \cdot \left|_{(\alpha)}^{dif}E_x\right|^2, & |z| \leq 2\pi\delta \end{cases}$, $\alpha = 3\pi\chi_{xxxx}^{(3)}$, see Fig. 1. The complete

diffraction field ${}^{dif}_{(\alpha)}E_x(y,z) = {}^{inc}E_x(y,z) + {}^{scat}_{(\alpha)}E_x(y,z)$ (here ${}^{scat}_{(\alpha)}E_x(y,z)$ scattering field) satisfies to conditions of a problem, see (4):

$$\left[\frac{\partial^2}{\partial y^2} + \frac{\partial^2}{\partial z^2} + \kappa^2{}_{(\alpha)}\varepsilon(z, |{}^{dif}_{(\alpha)}E_x|^2)\right]{}^{dif}_{(\alpha)}E_x(y,z) = 0, \qquad (15)$$

the generalized boundary conditions:

$${}^{dif}_{(\alpha)}E_{tg} \text{ and } {}^{dif}_{(\alpha)}H_{tg} \text{ are continuous at discontinuities } {}_{(\alpha)}\varepsilon(z, |{}^{dif}_{(\alpha)}E_x|^2);$$
$${}^{dif}_{(\alpha)}E_x(y,z) = {}^{dif}_{(\alpha)}U(z,\kappa,\phi)\cdot\exp(i\phi y) \text{ - condition of spatial quasi-homogeneity on } y; \qquad (16)$$

condition of the radiation for scattered field

$${}^{scat}_{(\alpha)}E_x(y,z) = \begin{Bmatrix} {}^{scat}_{(\alpha)}a \\ {}^{scat}_{(\alpha)}b \end{Bmatrix} \exp[i(\phi y \pm \Gamma(\kappa,\phi)\cdot(z \mp 2\pi\delta))], \quad z \begin{matrix}>\\<\end{matrix} \pm 2\pi\delta, \qquad (17)$$

satisfying to the requirement of absence of waves coming from infinity (8).

Here frequency $\kappa \in R$; $\kappa = \omega/c \equiv 2\pi/\lambda$; $c = (\varepsilon_0\mu_0)^{-1/2}$, $\varepsilon_0$, $\mu_0$ and $\lambda$ length of the wave is the parameters of environment; $\phi \equiv \kappa\cdot\sin(\varphi) \in R$, where $\varphi$ angle of fall of a plane wave ${}^{inc}E_x(y,z)$ which is counted in area $z > 2\pi\delta$ from normal to a layer against a course of the hour arrow $|\varphi| < \pi/2$ (see Fig. 1); amplitudes of an incident diffraction field ${}^{inc}a$ are given.

The required solution of a problem (15)-(17) has a kind:

$${}^{dif}_{(\alpha)}E_x(y,z) = {}^{dif}_{(\alpha)}U(z)\cdot\exp(i\phi y) =$$
$$= \begin{cases} {}^{inc}a\cdot\exp(i\cdot(\phi y - \Gamma\cdot(z-2\pi\delta))) + {}^{scat}_{(\alpha)}a\cdot\exp(i\cdot(\phi y + \Gamma\cdot(z-2\pi\delta))), & z > 2\pi\delta, \\ {}^{scat}_{(\alpha)}U(z)\cdot\exp(i\cdot\phi y), & |z| \leq 2\pi\delta, \\ {}^{scat}_{(\alpha)}b\cdot\exp(i\cdot(\phi y - \Gamma\cdot(z+2\pi\delta))), & z < -2\pi\delta. \end{cases} \qquad (18)$$

Here ${}^{dif}_{(\alpha)}U(-2\pi\delta) = {}^{scat}_{(\alpha)}b$, ${}^{dif}_{(\alpha)}U(2\pi\delta) = {}^{inc}a + {}^{scat}_{(\alpha)}a$.

The nonlinear problem (15)-(17) is reduced to finding the solutions ${}^{dif}_{(\alpha)}U(z) \in L_2([-2\pi\delta, 2\pi\delta])$ (see (18)) of the non-homogeneous nonlinear integrated equation of the second kind [4-7]:

$${}^{dif}_{(\alpha)}U(z) + \frac{i\kappa^2}{2\Gamma}\int_{-2\pi\delta}^{2\pi\delta}\exp(i\Gamma\cdot|z-z_0|)\cdot\left[1 - \left(\varepsilon^{(L)}(z_0) + \alpha|{}^{dif}_{(\alpha)}U(z_0)|^2\right)\right]\cdot{}^{dif}_{(\alpha)}U(z_0)\,dz_0 = {}^{inc}U(z), \quad |z| \leq 2\pi\delta, \qquad (19)$$

where ${}^{inc}U(z) = {}^{inc}a\exp[-i\Gamma\cdot(z-2\pi\delta)]$.

Similarly to results of the analysis of a linear diffraction problem, the nonlinear diffraction problems (15)-(17) with use (18) are reduced to a finding of the solutions of the non-homogeneous system of the nonlinear equations of the second kind:

$$(E -_{(\alpha)} B(\kappa, \phi, |_{(\alpha)}^{dif} U|^2)) \cdot_{(\alpha)}^{dif} U = {}^{inc} U \qquad (20)$$

where, matrix of system (20) for a required field $_{(\alpha)}^{dif} U$ is set similarly to matrix of system (14) for $_{(\alpha)}^{(nor)} U$ (see (14)). Here $_{(\alpha)}^{dif} U = \{_{(\alpha)}^{dif} U_n\}_{n=1}^{N}$ - vector-column of unknown $_{(\alpha)}^{dif} U_n = _{(\alpha)}^{dif} U(z_n, \kappa, \phi)$, given in units, $z_1 = -2\pi\delta < z_2 < ... < z_n < ... < z_N = 2\pi\delta$, $n = 1,2,...,N$; $N$ - number of units, determining the order of system (20); $E = \{\delta_{n\,m} \equiv \delta_n^m\}_{n,m=1}^{N}$ (where $\delta_n^m$ - Kronecker delta) and $_{(\alpha)} B\left(\kappa, \phi, |_{(\alpha)}^{dif} U|^2\right) = \left\{A_m \cdot_{(\alpha)} K_{n\,m}\left(\kappa, \phi, |_{(\alpha)}^{dif} U|^2\right)\right\}_{n,m=1}^{N}$ - matrix of dimension $N \times N$; $_{(\alpha)} K_{n\,m}\left(\kappa, \phi, |_{(\alpha)}^{dif} U|^2\right) = \frac{i\kappa^2}{2\Gamma(\kappa, \phi)} \cdot \exp(i\,\Gamma(\kappa, \phi) \cdot |z_n - z_m|) \cdot \left[1 - \left(\varepsilon^{(L)}(z_m) + \alpha \cdot |_{(\alpha)}^{dif} U_m|^2\right)\right]$; $A_m$ - numerical coefficients dictated by chosen quadrature form. The vector-column of the right hand part of (20) is the given current ${}^{inc} U = \{{}^{inc} U(z_n) \equiv {}^{inc} a \cdot \exp[-i \cdot \Gamma \cdot (z_n - 2\pi\delta)]\}_{n=1}^{N}$.

Solutions of non-homogeneous nonlinear system of the equations (20) are carried out by the method of iterations [10-15]. First step can be a finding of the solution of a linear problem (case $\alpha = 0$) equivalent diffraction problem of a flat wave ${}^{inc} U$ on a linear layer $_{(\alpha=0)} \varepsilon\left(z, |_{(\alpha=0)}^{dif} E_x|^2\right) \equiv \varepsilon^{(L)}(z)$. Each subsequent step of approximation represents the solution of linear diffraction problem received in result of linearization of the initial nonlinear system (20) for a nonlinear layer $_{(\alpha)} \varepsilon\left(z, |_{(\alpha)}^{dif} E_x|^2\right)$. This process shall present as:

$$\left\{\left(E -_{(\alpha)} B\left(\kappa, \phi, |_{(\alpha)}^{dif\,(m-1)} U|^2\right)\right) \cdot_{(\alpha)}^{dif\,(m)} U = {}^{inc} U\right\}_{m=1}^{M : \left\|_{(\alpha)}^{dif\,(m)} U -_{(\alpha)}^{dif\,(m-1)} U\right\| / \left\|_{(\alpha)}^{dif\,(m)} U\right\| < \xi_1} \qquad (21)$$

Here index $(m)$ - designates the step of iteration, $M$ - number of a step, $M : \left\|_{(\alpha)}^{dif\,(m)} U -_{(\alpha)}^{dif\,(m-1)} U\right\| / \left\|_{(\alpha)}^{dif\,(m)} U\right\| < \xi_1$ is the condition of the ending of iterative process, $\|\cdot\|$ - norm in space of the solutions $_{(\alpha)}^{dif\,(m)} U$, $\xi_1$ - given meaning of a relative error.

Other algorithms of the solution of the nonlinear integrated equation (19) on the basis of a method of Newton with use of Simpson's quadrature and a method of Newton with Taylor's decomposition of a field of diffraction contain in [10-15].

## 3.2 Effects of Resonant Scattering of the Intensive Fields

### 3.2.1 Intensity and Resonant Frequency

The effect of non-uniform shift of resonant frequency of the diffraction characteristics of nonlinear dielectric layer is found out at increase of intensity of inciting field [13-17] (see Fig. 2, and also Fig. 3). Growth of intensity of the inciting field $|I| = |{}^{inc} a|$ results in change $\eta\left(|R(\alpha)|^2\right) = |R(\alpha)|^2 / |I|^2$: reduction of value of resonant frequency with increase and reduction of a steepness of the diffraction characteristics before and after resonant frequency (Fig. 2). Here: $|R(\alpha)| \equiv \left|{}^{scat}_{(\alpha)}a\right|$, $|T(\alpha)| \equiv \left|{}^{scat}_{(\alpha)}b\right|$ and $|I|^2 = |T(\alpha)|^2 + |R(\alpha)|^2$.

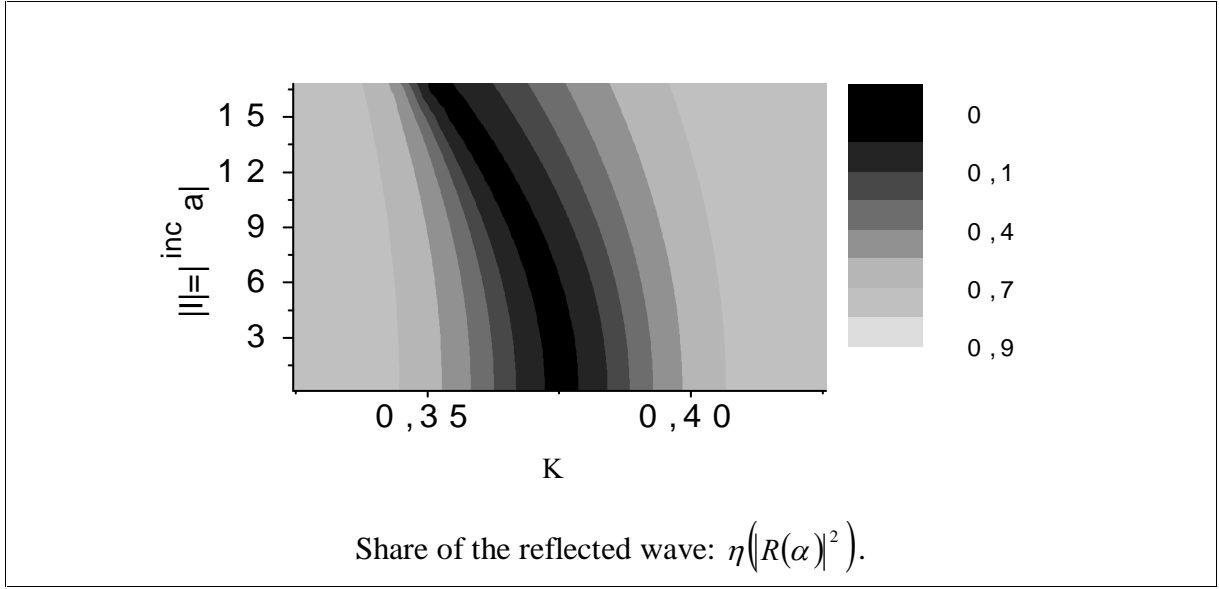

Share of the reflected wave: $\eta\left(|R(\alpha)|^2\right)$.

**Figure 2.** Parameters of the nonlinear problem:
$\alpha = 0{,}01$; $\varepsilon^{(L)} = 16$; $\delta = 0{,}5$; $\varphi = 45^0$.

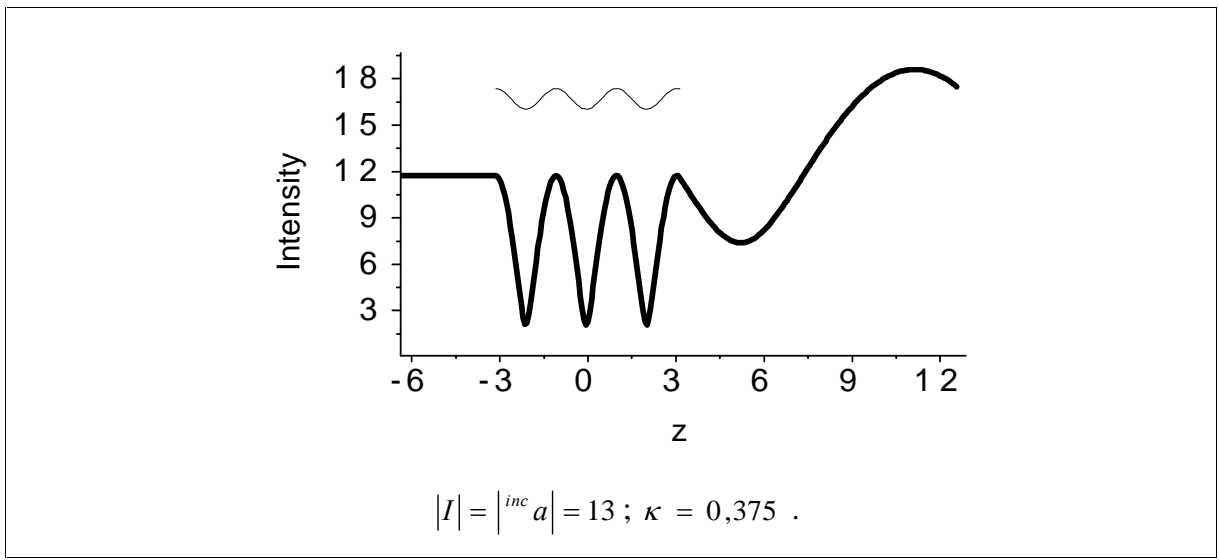

$|I| = \left|{}^{inc}a\right| = 13$; $\kappa = 0{,}375$.

**Figure 3.** Parameters of structure and designations:
$\alpha = 0{,}01$; $\varepsilon^{(L)} = 16$; $\delta = 0{,}5$; $\varphi = 45^0$;
——— $\left|{}^{dif}_{(\alpha)}E_x\right|$;  ———  ${}_{(\alpha)}\varepsilon\left(z, \left|{}^{dif}_{(\alpha)}E_x\right|^2\right)$.

### 3.2.2 Intensity and Transparency

The effect of increase of the angle of the transparency of the nonlinear layer ($\alpha \neq 0$) at growth of intensity of the inciting field is found out, [13-17]. See Fig. 4 and Fig. 5, A: $|{}^{inc}a| = 8$, $\varphi \approx 46^0$ and B: $|{}^{inc}a| = 11$, $\varphi \approx 74^0$.

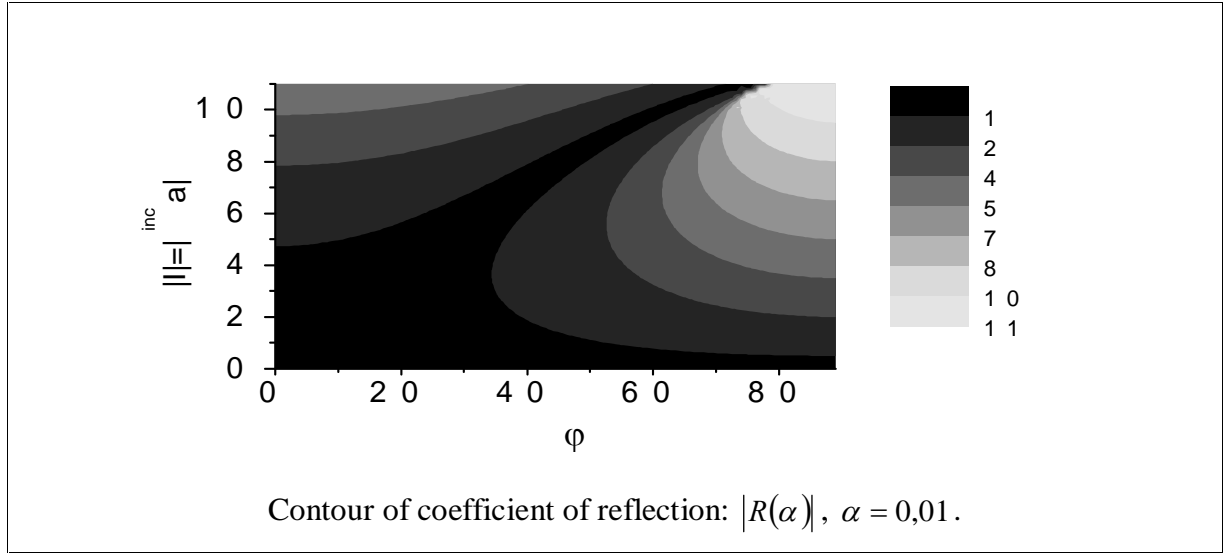

Contour of coefficient of reflection: $|R(\alpha)|$, $\alpha = 0,01$.

**Figure 4.** Parameters of the nonlinear problem: $\varepsilon^{(L)} = 16$; $\delta = 0,5$; $\kappa = 0,375$.

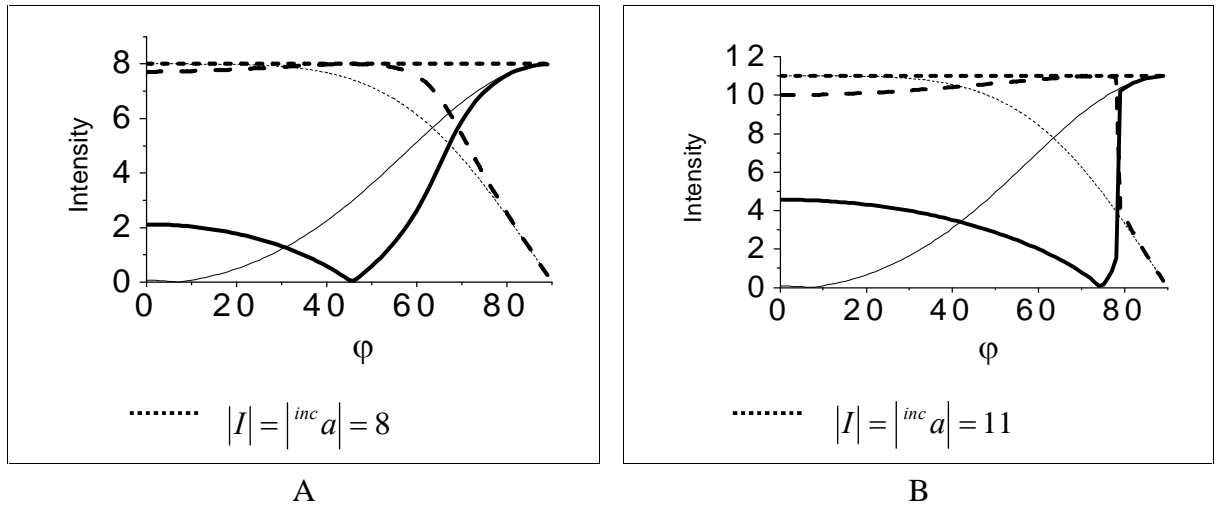

A    B

**Figure 5.** Parameters and designations for nonlinear ($\alpha = 0,01$) and linear ($\alpha \equiv 0$) layer:

$$\varepsilon^{(L)} = 16;\ \delta = 0,5;\ \kappa = 0,375;$$

— $|R(\alpha)|$ and - - - $|T(\alpha)|$ for **nonlinear layer** with $\alpha = 0,01$;

— $|R(\alpha)|$ and ----- $|T(\alpha)|$ for **linear layer** with $\alpha \equiv 0$.

These effects (see sections 3.2.1 and 3.2.2) are connected to resonant properties of a nonlinear dielectric layer and caused by increase of a variation of dielectric permeability of a layer (its nonlinear components) at increase of intensity of a field of excitation of researched nonlinear object, see Fig. 3.

The given results of calculations are received with use of the iteration scheme (21). In a considered range of a variation of parameters of a nonlinear problem of diffraction systems of the equations of dimension were used $N = 101$. Also the relative size of an error was set $\xi_1 = 10^{-7}$.

## 4 The Algorithm of Calculation is the Amplitudes-Phase Dispersion of Eigen Oscillation-Wave Fields of Nonlinear Layer. Norm of Eigen Fields of Nonlinear Structure.

The algorithm of numerical definition of the untrivial solutions of homogeneous nonlinear system of the equations (14) is carried out by the method of successive approximations. The norm of eigen field is defined from the solution of a diffraction problem (20) of a nonlinear layer [18-20].

On a first step we find initial approximation of iterative process. We solve a linear spectral problem (14) and diffraction problem (20) when $\alpha = 0$, equivalent linear spectral problem (5)-(7) and linear diffraction problem (15)-(17) for a layer $_{(\alpha=0)}\varepsilon\left(z, \left|_{(\alpha=0)}^{dif}E_x\right|^2\right) \equiv \varepsilon^{(L)}(z)$. Each subsequent step of approximation represents the solution of the linear spectral problem received in result of linearization of the initial nonlinear system (14) for a nonlinear layer $_{(\alpha)}\varepsilon\left(z, \left|_{(\alpha)}^{(nor)}E_x\right|^2\right)$. In the field of space engaged by nonlinear object the principle of a superposition is not carried out. Therefore, but step ($s$) of iteration, norm $_{(\alpha)}^{(nor)}a^{(s-1)} \equiv _{(\alpha)}a^{scat\,(M)}$ (or $_{(\alpha)}^{(nor)}b^{(s-1)} \equiv _{(\alpha)}b^{scat\,(M)}$) (see (12) and (18)) of an eigen field of a nonlinear layer is determined from the solution of a nonlinear diffraction problem (15)-(17), by use of iterative process (21) for $\kappa = \mathrm{Re}\,\kappa$ and $\phi = \mathrm{Re}\,\phi$. Schematically this process shall present as:

$$\left\{\begin{array}{l} \text{The Nonlinear Diffraction Problem:} \\ \left\{\left(E - _{(\alpha)}B\left(\mathrm{Re}\left(\kappa^{(s-1)}\right), \mathrm{Re}\left(\phi^{(s-1)}\right), \left|_{(\alpha)}^{dif\,(m-1)}U\right|^2\right)\right) \cdot _{(\alpha)}^{dif\,(m)}U = {^{inc}U}\right\}_{m=1}^{M:\left\|_{(\alpha)}^{dif\,(m)}U - _{(\alpha)}^{dif\,(m-1)}U\right\| / \left\|_{(\alpha)}^{dif\,(m)}U\right\| < \xi_1} \\ \text{The Norm Problem:} \\ _{(\alpha)}^{(nor)}a^{(s-1)} \equiv _{(\alpha)}a^{scat\,(M)} \quad \left(or \quad _{(\alpha)}^{(nor)}b^{(s-1)} \equiv _{(\alpha)}b^{scat\,(M)}\right). \\ \text{The Spectral Problem:} \\ \left\{_{(\alpha)}f\left(\kappa^{(s)}, \phi^{(s)}, \left|_{(\alpha)}^{(nor)}U^{(s-1)}\right|^2\right) = \det\left(E - _{(\alpha)}B\left(\kappa^{(s)}, \phi^{(s)}, \left|_{(\alpha)}^{(nor)}U^{(s-1)}\right|^2\right)\right) = 0,\right. \\ (E - _{(\alpha)}B(\kappa^{(s)}, \phi^{(s)}, \left|_{(\alpha)}^{(nor)}U^{(s-1)}\right|^2))_{(\alpha)}^{(nor)}U^{(s)} = 0, \\ here: \kappa^{(s)} \in _{(\alpha)}\Omega_\kappa^{(s)}; \phi^{(s)} \in _{(\alpha)}\Omega_\phi^{(s)}; (\kappa^{(s)}, \phi^{(s)}) \in _{(\alpha)}\Omega_{(\kappa,\phi)}^{(s)}. \end{array}\right\}_{s=1}^{S} \quad (12)$$

Here index $(s)$ designates the step of iteration, $S$ - number of a step, $S: \left\| {}^{(nor)}_{(\alpha)}U^{(s)} - {}^{(nor)}_{(\alpha)}U^{(s-1)} \right\| / \left\| {}^{(nor)}_{(\alpha)}U^{(s)} \right\| < \xi$ is the condition of the ending of iterative process (22), $\|\cdot\|$ - one of norms in space of the solutions ${}^{(nor)}_{(\alpha)}U^{(s)}$, $\xi$ is given meaning of a relative error.

## 5 The Further Researches

The offered algorithm of the solution of nonlinear spectral problems underlies creations of the constructive approach of the local description of processes of the space-time evolutions of the electromagnetic field of the open nonlinear electrodynamics structures.

The further researches are connected to development of the approach of the description of evolutionary processes near to critical points of the amplitude-phase dispersion of nonlinear structure. The case of a linear problem in [6], [21] (see also Tab. 1) is considered.

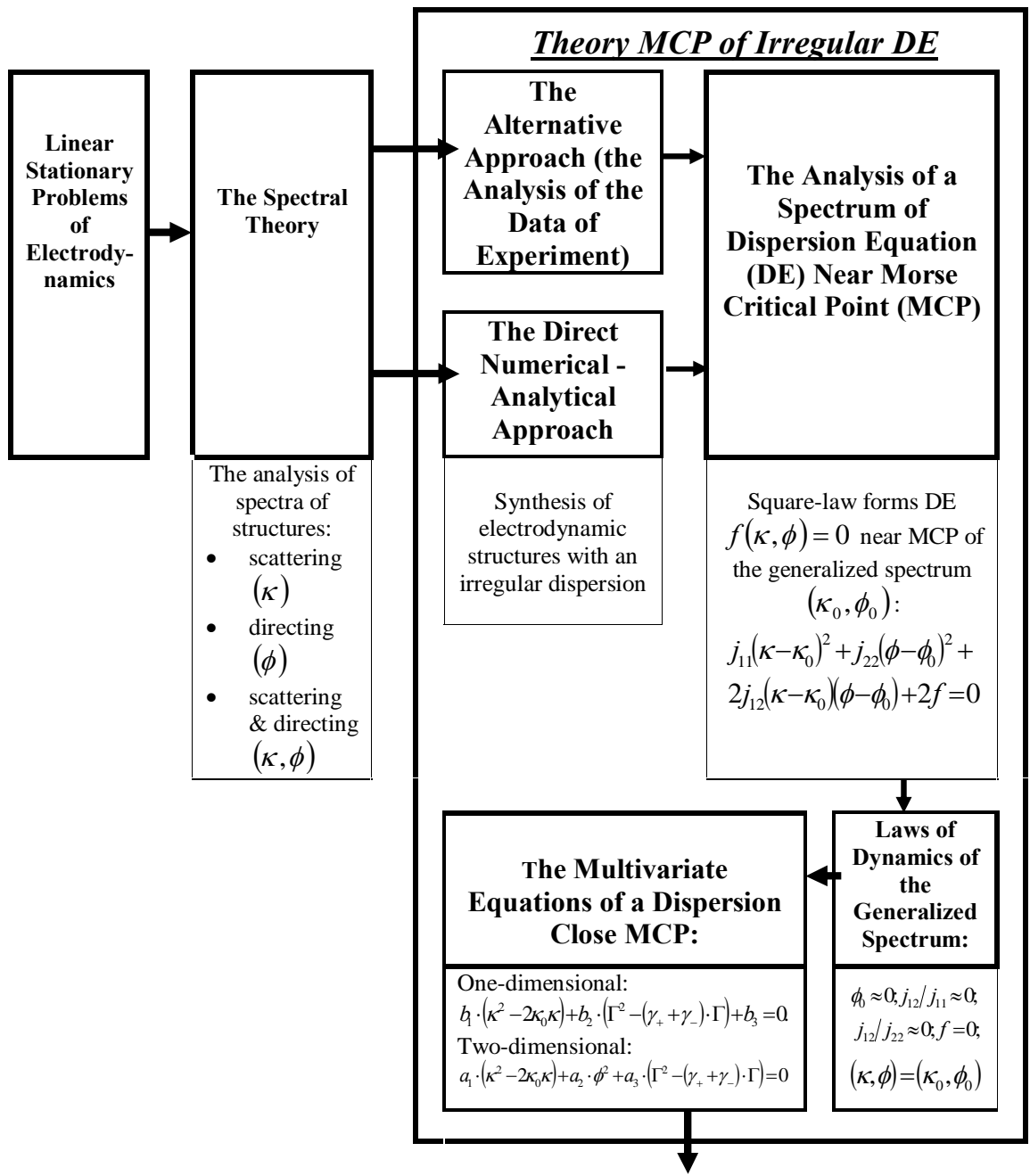

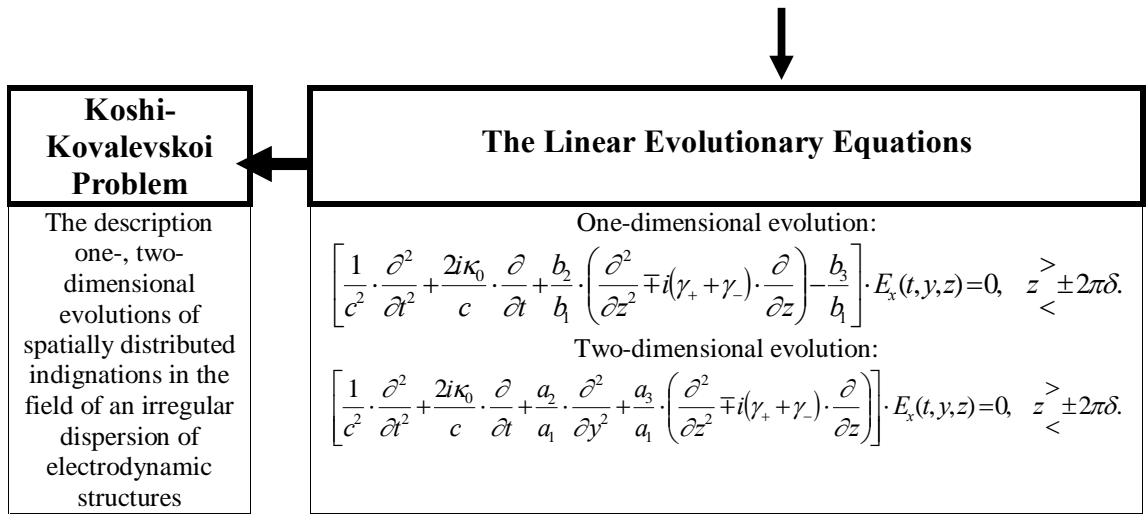

**Table 1.** Dispersion and evolution processes near to critical points for the linear problem.

# 6 Conclusion

The proposed algorithms and results of the numerical analysis are applied: at the analysis of amplitude-phase dispersion of eigen oscillation-wave fields in the nonlinear objects [18-20]; development of the constructive approach of the description of evolutionary processes near to critical points of the amplitude-phase dispersion of nonlinear structure (the case of a linear problem in [6], [21] is considered) and analysis of the evolution of field [22] ; at investigation of processes of wave self-influence [2]; at designing new selecting energy; transmitting, remembering devices; etc.